\documentclass[aps,prl,superscriptaddress,showpacs,twocolumn,nofootinbib,longbibliography]{revtex4-1}
\usepackage[pdftex]{graphicx}
\usepackage{amsmath}
\usepackage{xspace}
\usepackage{color}
\usepackage[colorlinks=true,linkcolor=blue,urlcolor=blue,citecolor=blue]{hyperref}
\usepackage[normalem]{ulem}
\usepackage{mathtools}
\usepackage{bm,bbold,dsfont,amsfonts,mathrsfs,amssymb}

\usepackage{enumitem}

\usepackage{textcomp}

\usepackage{hyperref}
\newcounter{myequation}
\makeatletter
\@addtoreset{equation}{myequation}
\makeatother

\DeclareMathOperator{\tr}{\mbox{tr}}

\newcommand{\ii}{{\mathbf i}}
\newcommand{\jj}{{\mathbf j}}

\usepackage{color}
\definecolor{BV}{rgb}{0.1,0.,0.6}
\definecolor{R}{rgb}{0.9,0,0}
\definecolor{G}{rgb}{0.2,0.8,0.2}
\usepackage{comment}

\renewcommand{\L}{{\mathcal L}}

\newcommand{\be}{\begin{equation}}
\newcommand{\ee}{\end{equation}}

\begin{document}

\title{
Quantum dynamics in one and two dimensions  via the recursion method
 }

	\author{Filipp Uskov}
%	\email{}
	\affiliation{%
		Skolkovo Institute of Science and Technology\\
		Bolshoy Boulevard 30, bld. 1, Moscow 121205, Russia
	}
    \affiliation{Gubkin Russian State University of Oil and Gas\\
    65 Leninsky Prospekt, Moscow 119991, Russia
	}

	\author{Oleg Lychkovskiy}
\email{o.lychkovskiy@skoltech.ru}
	\affiliation{%
		Skolkovo Institute of Science and Technology\\
		Bolshoy Boulevard 30, bld. 1, Moscow 121205, Russia
	}%

	\date{\today}

	%%%%%%%%%%%%%%%%%%%%%%%%%%%%%%%%%%%%%%%%%
	%%%%%%%%%%%%%%%%%%%%%%%%%%%%%%%%%%%%%%%%%
\begin{abstract}
We report an implementation of the recursion method that  addresses quantum many-body dynamics in the nonperturbative regime. The method essentially amounts to constructing a Lanczos basis in the space of operators and solving coupled Heisenberg equations in this basis.  The reported implementation has two key ingredients: a computer-algebraic routine for symbolic calculation of nested commutators and a procedure to extrapolate the sequence of Lanczos coefficients according to the universal operator growth hypothesis. We apply the method to calculate infinite-temperature correlation functions for spin-$1/2$ systems on one- and two-dimensional lattices. In two dimensions the  accessible  timescale is large enough to essentially embrace the relaxation to equilibrium.  The method allows one to accurately calculate transport coefficients.  As an illustration, we compute the diffusion constant for the transverse-field Ising model on a square lattice.
\end{abstract}

	\pacs{}

	\maketitle

\noindent{\it Introduction.}~ Quantum dynamics is one of the central topics in condensed matter physics. While for one-dimensional (1D) systems various numerical approaches typically deliver highly satisfactory results, addressing higher dimensions turns out to be much more challenging. Diverse techniques are being developed to tackle quantum dynamics in two and three dimensions, including determinant quantum Monte-Carlo \cite{Huang_2019_Strange}, MPS computations on infinite cylinders \cite{James_2015_Quantum,Gohlke_2017_Dynamics,Hashizume_2022_Dynamical,DeNicola_2022_Entanglement}, methods based on  projected entangled pair states \cite{Dziarmaga_2021_Time,kaneko2023dynamics},  functional renormalization group \cite{Krieg_2019_Exact,Tarasevych_2021_Dissipative}, classical approximations~\cite{Schubert_2021_Quantum}, hybrid quantum-classical methods  \cite{Starkov_2018_Hybrid,Starkov_2020_Free},  unfolding of two-dimensional (2D) to nonlocally coupled  1D  systems \cite{thomson2023unraveling} {\it etc}.

Here we report an implementation of the recursion method capable of addressing high-temperature dynamics of 1D and 2D lattice systems. The recursion method has a long history \cite{viswanath2008recursion,Haydock_1980_Recursive}, however instances of its application to many-body systems are relatively scarce \cite{Jensen_1973_Sixth,Engelsberg_1975_Approximants,Sen_1991_Transition,Cai_1992_Long-time,Viswanath_1994_Ordering,Bohm_1994_Spin,Lindner_2014_Conductivity,Khait_2016_Spin,Auerbach_2018_Hall,Parker_2019,deSouza_2020_Dynamics,Yates_2020_Lifetime,Yates_2020_Dynamics,Yuan_2021_Spin}. The basic object of the recursion method is a sequence of Lanczos coefficients $b_n$, $n=0,1,2,\dots$, that are to be obtained from the nested commutators of the system Hamiltonian with the observable in question. This sequence becomes infinite in the thermodynamic limit. At the same time, the complexity  of calculating the Lanczos coefficients  grows factorially with $n$. This has been hindering the application of the method for decades.

We alleviate the above difficulty by two complementary remedies. First, we develop a computer algebra routine to calculate a record number of nested commutators. The computation is performed directly in the thermodynamic limit and keeps the Hamiltonian parameters symbolic. Second, we extrapolate the remaining part of the Lanczos sequence according to the universal operator growth hypothesis (UOGH) \cite{Parker_2019} and other recent insights in the asymptotic behaviour of this sequence \cite{Yates_2020_Lifetime,Yates_2020_Dynamics,Bhattacharjee_2022_Krylov,avdoshkin2022krylov}. Remarkably, the extrapolation works better the further the system is from integrable points. This makes our approach inherently nonperturbative.

The paper is organized as follows. We start from introducing basic concepts and definitions, in particular, the autocorrelation function. Then we discuss the truncated Taylor expansion of the autocorrelation function. After that we outline the recursion method, the UOGH, and a procedure to obtain transport coefficients from the Lanczos sequence. Next we describe our implementation of the recursion method. Then the method is applied to one 1D model and two 2D models. Discussion and outlook conclude the paper.

%\cite{Engelsberg_1974} FID in CaF2 and refs to analytical moments of this 3D material.

%\cite{Engelsberg_1975} simple instance of recursion method for FID in CaF2.

%\cite{Elsayed_2014_Signatures} is a precursor work to UOGH

%Up to 8'th moment for 3D dipole interaction \cite{Jensen_1973_Sixth}

%2D Ising with MPS and ED ($5\times 5$ lattice) \cite{kaneko2023dynamics}

%Up to $\mu_{12}$ for 2D hard-core bosons \cite{Lindner_2014_Conductivity}

%%%%%%%%%%%%%%%%%%%%%%%%%%%%%%%%%%%%%%%%%%%%%%%%%%%%%%%%%%%%%%%%%%%%%%%%%%%%%%%%%%%%%%%%%%%%%%
%%%%%%%%%%%%%%%%%%%%%%%%%%%%%%%%%%%%%%%%%%%%%%%%%%%%%%%%%%%%%%%%%%%%%%%%%%%%%%%%%%%%%%%%%%%%%%

\medskip
\noindent{\it Autocorrelation function.}~
We consider a quantum system with a Hamiltonian $H$ and focus on some observable given by a self-ajoint Shr\"odinger operator $A$. The same observable in the Heisenberg representation reads $A(t)=e^{i t H} A \, e^{-i t H}$. It is convenient to introduce the commutation superoperator $\L\equiv [H,\bullet]$. Then the Heisenberg equation of motion reads $\partial_t A(t)=i\, \L A(t)$, and its formal solutionis is given by $A(t)=e^{i\,t\,\L} A$.

Throughout the paper we focus on the normalized infinite-temperature autocorrelation function
\begin{equation}\label{autocorrelation function}
C(t)\equiv \tr\big( A(t) A \big)/\tr A^2.
\end{equation}
It has the properties $C(0)=1$ and $C(-t)=C(t)$. We remark that strong long-lived quantum correlations can well exist at infinite temperature \cite{Kanasz-Nagy_2017_Quantum}.

It is convenient to introduce a scalar product in the space of operators according to
\begin{equation}\label{scalar product}
\big(A|B\big)\equiv \tr \big(A^\dagger B\big)/d,
\end{equation}
where $d$ is the Hilbert space dimension (which is assumed to be finite). The scalar product entails the norm $\|A\|=\sqrt{(A|A)}$.  In this notations, the autocorrelation function can be written as $C(t)=(A(t)|A)/\|A\|^2$. The superoperator $\L$ is self-adjoint with respect to this scalar product.

\medskip
\noindent{\it Truncated Taylor expansion.}~
Expanding $A(t)$ in powers of $t$, one obtains the Taylor expansion of the autocorrelation function,
\begin{equation}\label{Taylor expansion}
C(t) \equiv \sum_{m=0}^\infty (-1)^m \, \frac{\mu_{2m}}{(2m)!} \, t^{2m},
\end{equation}
with even moments given by
\begin{equation}\label{moments}
\mu_{2m}\equiv(\L^{2m} A|  A)/\|A\|^2=(\L^m A| \L^m A)/\|A\|^2
\end{equation}
and odd moments being zero, ensuring that the autocorrelation function is even. Note that $\mu_0=1$ by definition.

The Taylor expansion \eqref{Taylor expansion} is known to have an infinite convergence radius for 1D systems with short-range interactions \cite{Araki_1969_Gibbs} and a finite convergence radius in higher dimensions \cite{Parker_2019}.

\begin{figure*}[t] %  figure placement: here, top, bottom, or page
		\centering
\includegraphics[width=0.95\linewidth]{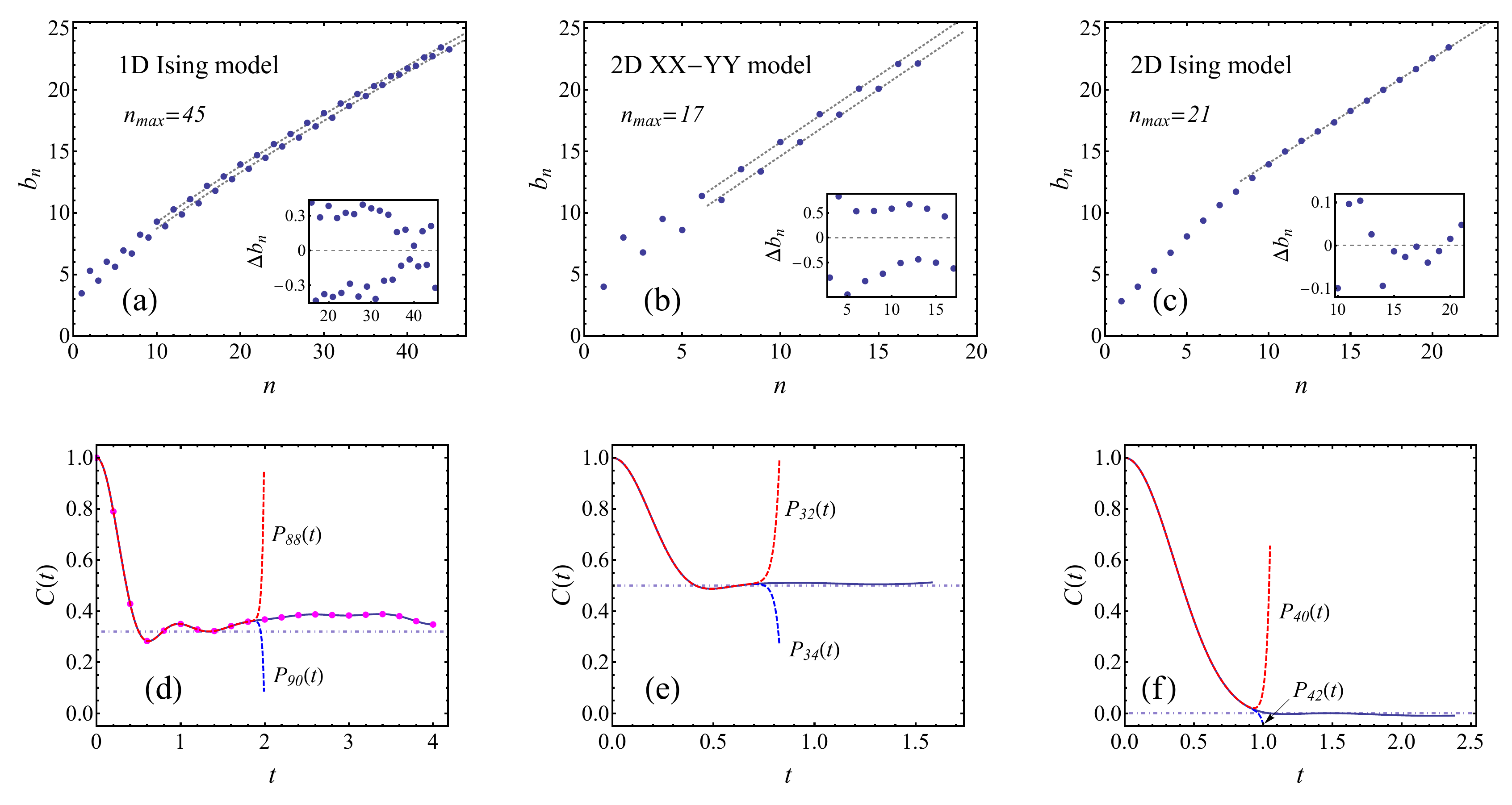}
        \caption{Upper row: Lanczos coefficients for three models considered in the text. Dashed lines indicate extrapolating functions~\eqref{b_n extrapolation 1D},\eqref{b_n extrapolation 2D}. Insets highlight the subleading contribution $\Delta b_n$, where $\Delta b_n=b_n-(\alpha \, n/\log \,n+\gamma)$ in 1D and $\Delta b_n=b_n-(\alpha \, n +\gamma)$ in~2D. Lower row: correlation functions for the same models (solid lines) plotted up to $t=t_{\max}$. Dashed lines -- upper and lower polynomial bounds \eqref{inequality}. Horizontal dash-dotted lines - long-time averages $\overline C$. The result for the 1D Ising model is benchmarked by the exact diagonalization (dots).
		\label{fig 1}
}
\end{figure*}

Truncating the Taylor expansion \eqref{Taylor expansion} at the order $2n$, one obtains a polynomial $P_{2n}(t)$. Remarkably, these polynomials constitute rigorous upper and lower bounds on the autocorrelation function \cite{Platz_1973_Rigorous,Roldan_1986_Dynamic,Brandt_1986_High,Bohm_1992_Dynamic},
\begin{equation}\label{inequality}
P_{4l+2}(t) \leq C(t) \leq P_{4l}(t),\quad l=1,2,\dots
\end{equation}
These two-sided bounds are extremely tight up to a certain  time,  allowing one to precisely benchmark  more sophisticated approximations to $C(t)$, see Fig. \ref{fig 1}.

\medskip
\noindent{\it Recursion method.}~ We employ the Heisenberg-picture version of the recursion method \cite{viswanath2008recursion}. It is essentially about solving coupled Heisenberg equations in the orthogonal Lanczos  basis $\{ A_n\},\, n=0,1,2,\dots$ defined iteratively as follows: $|A_0)=\|A\|^{-1}\,|A)$, $|A_1)=\L |A_0)$,
\begin{align}\label{Lanczos basis}
b_n & = \|A_n\|,\quad n=0,1,2,\dots,\nonumber\\
|A_n) & =b_{n-1}^{-1} \, \L |A_{n-1}) - b_{n-1}\,b_{n-2}^{-1}  |A_{n-2}) \quad n=2,3,\dots
\end{align}
The superoperator $\L$ acquires a tridiagonal form in this basis, with the zero main diagonal and the sequence of Lanczos coefficients $b_n$ in the sub/supra-diagonals. As a result, the autocorrelation function \eqref{autocorrelation function} enters a set of coupled equations
\begin{align} \label{coupled equations}
  \partial_t \varphi_n (t) & = -b_{n+1} \, \varphi_{n+1}(t) + b_n  \, \varphi_{n-1} (t),\quad n=0,1,2,\dots \nonumber\\
  C(t) & =\varphi_0(t),
\end{align}
where  $\varphi_{-1}(t)\equiv 0$ and $\varphi_n (0) = \delta_{0n}$.

This way the autocorrelation function becomes implicitly determined by the sequence of Lanczos coefficients~$b_n$. These coefficients can be obtained recurrently according to \eqref{Lanczos basis} or, alternatively, from the moments~\eqref{moments}~\cite{Sanchez-Dehesa_1978_Spectrum}.

\medskip
\noindent{\it UOGH and extrapolation of Lanczos coefficients.}~In practice, only a finite number of $b_n$ can be computed. Other coefficients are to be extrapolated. The UOGH put forward in~\cite{Parker_2019}  states that for generic systems the leading asymptotics of $b_n$ is linear, with a logarithmic correction in one dimension (see also an earlier paper \cite{Elsayed_2014_Signatures} for a similar result for classical systems). It has been further revealed that certain subleading terms of the asymptotics can be equally important for the dynamics \cite{Viswanath_1994_Ordering,Parker_2019,Yates_2020_Lifetime,Yates_2020_Dynamics,Dymarsky_2021_Krylov,Bhattacharjee_2022_Krylov,avdoshkin2022krylov,Camargo_2023_Krylov}. Guided by these insights, we employ the following extrapolation formulae for $n \gg 1$:
\begin{align}
  b_n &\simeq \alpha n/\log n + \gamma + (-1)^n \gamma_*  & {\rm for}~~1{\rm D},  \label{b_n extrapolation 1D}\\
  b_n & \simeq  \alpha n + \gamma + (-1)^n \gamma_*  & {\rm for}~~2{\rm D}, \label{b_n extrapolation 2D}
\end{align}
Here $\alpha$, $\gamma$ and $\gamma_*$ are the fitting parameters.  In particular, $\gamma_*$ parameterizes  odd-even alterations in the Lanczos sequence that emerge whenever $\overline C \equiv \lim_{t\rightarrow \infty} C(t)$ is nonzero ({\it cf.} \cite{Viswanath_1994_Ordering,Yates_2020_Lifetime,Yates_2020_Dynamics}).

\medskip
\noindent{\it Transport coefficients.}~
Whenever $A=J$ is the current of some conserved quantity, the autocorrelation function of $J$ determines the corresponding transport coefficient~\cite{Kubo_1957_Statistical}. In particular, when the conserved quantity in question is energy, one can calculate the energy diffusion constant $D$ as~\cite{Kubo_1957_Statistical,Steinigeweg_2009_Density,Steinigeweg_2009_Density_for_spin_chains}
\begin{equation}\label{diffusion constant}
D=\frac{\|J\|^2}{\|H\|^2} \, \c{C},\qquad  \c{C} \equiv \int_0^\infty dt \,C(t),
\end{equation}
where $\tr H =0$ is assumed.

\begin{figure}[t] %  figure placement: here, top, bottom, or page
		\centering
\includegraphics[width=0.95\linewidth]{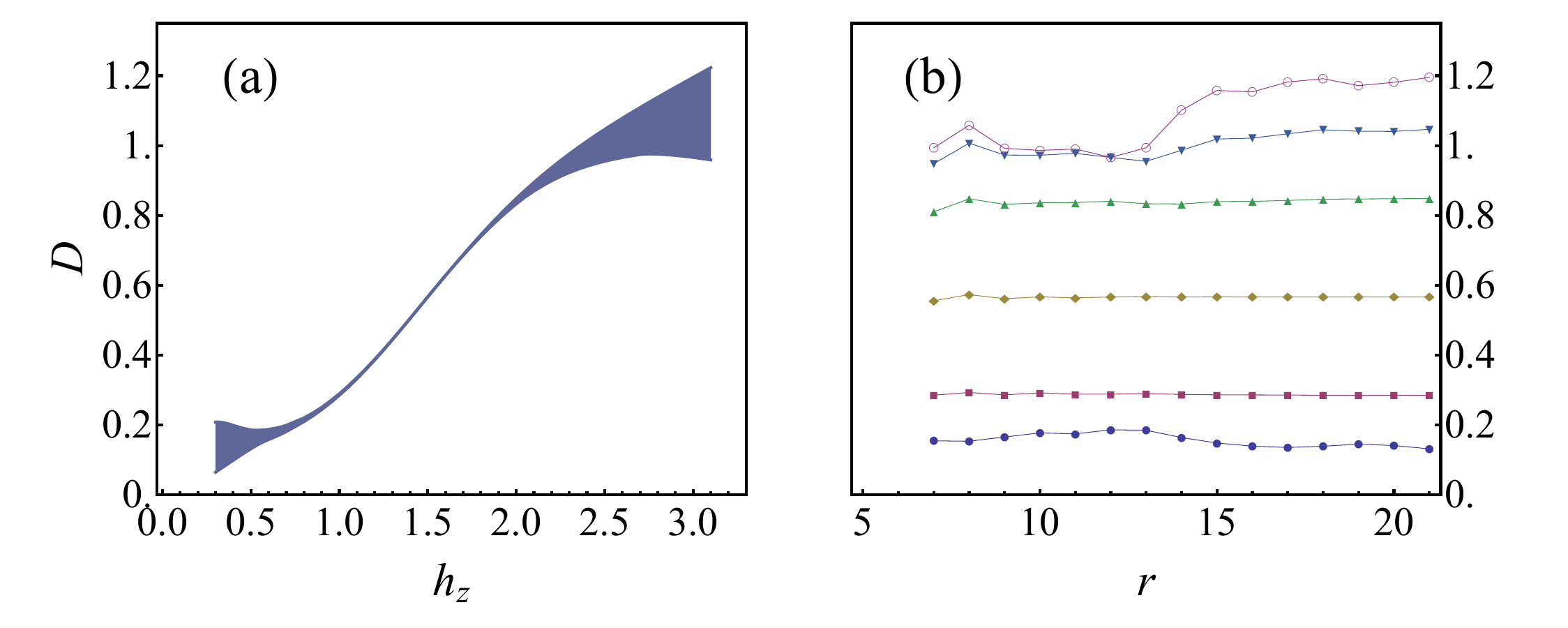}
        \caption{ (a) Diffusion constant  for the 2D Ising model \eqref{H 2D Ising} as a function of the transverse field $h_z$. The width of the line indicates the estimated uncertainty. (b)  Convergence of the diffusion constant with the approximation order $r$. Shown are data for fields $h_z=0.5, 1, 1.5, 2, 2.5, 3$ (from bottom to top).
		\label{fig 2}
}
\end{figure}

It has been shown recently that, employing the UOGH, one can obtain a precise approximation $\c{C}_r$ to  $\c{C}$  from a moderate number $r$ of known Lanczos coefficients \cite{wang2023diffusion}. The approximation reads \cite{wang2023diffusion} (see also \cite{bartsch2023equality,Joslin_1986_Calculation})
\begin{equation}\label{approximate C}
  \c{C}_r = \frac{1}{b_{r}}\prod_{m=1}^{[r/2]} \frac{b_{2m}^2}{b_{2m-1}^2}\times
      \begin{cases}
      1/p_{r}& \text{for even $r$,} \\
      p_{r}& \text{for odd $r$,}
    \end{cases}
\end{equation}
where $[r/2]$ is the integer value of $(r/2)$ and
\begin{equation}\label{p_r}
p_r=\Gamma\left(\frac{r}{2}+\frac{\gamma}{2\alpha}\right)\Gamma\left(\frac{r}{2}+\frac{\gamma}{2\alpha}+1\right)/
\left(\Gamma\left(\frac{r}{2}+\frac{\gamma}{2\alpha}+\frac12 \right)\right)^2.
\end{equation}
$ \c{C}_r$ usually converges to $ \c{C}$ rapidly upon increasing $r$ \cite{wang2023diffusion}, which is confirmed by our calculations, see Fig. \ref{fig 2}(b). We note that one can also substitute a truncated Taylor expansion in eq. \eqref{diffusion constant}, however the transport coefficients obtained this way are less accurate \cite{Morita_1972_Spin,Labrujere_1982_Spin,Cowan_1989_Spin,Tarasevych_2021_Dissipative}.

\medskip
\noindent{\it Symbolic implementation.}~
We consider one-dimensional chains and two-dimensional square lattices of spins $1/2$ with nearest-neighbour interactions. Both the Hamiltonian $H$ and the observable $A$ are considered to be translation-invariant.

The core routine of our method is a symbolic computation of nested commutators $\L^n A$. Importantly, the Hamiltonian parameters are also kept symbolic.  As compared to computation with numerical parameters, this requires essentially no overhead in terms of computational time and a moderate overhead in terms of memory. The major advantage of a fully symbolic calculation  is that it covers the whole parameter space in a single run.

%Upon commutation with a two-body Hamiltonian, the support of an operator is, in general, increased by one.
The computation is performed in the thermodynamic limit from the outset. The support of $\L^n A$ grows linearly with $n$, while the number of terms grows factorially. Since $\L$ is linear, the computation is straightforwardly parallelizable \cite{Brenes_2019_Massively}.  Computation of $\L^n A$ is the most resource-consuming routine of our code.

At the next step the moments \eqref{moments} are computed. They have the form of polynomials with respect to Hamiltonian parameters. For each model considered below, we list  several first moments in the text. The complete list of computed moments is available as a Supplementary Material~\cite{supp}.

Next we use the relation between Lanczos coefficients and moments \cite{Sanchez-Dehesa_1978_Spectrum} to compute $b_n$. At this step numerical values of the Hamiltonian parameters are plugged in. To avoid numerical instabilities, the rational arithmetics is used. As a result, a sequence of numerical Lanczos coefficients $b_n$, $n=0,1,\dots,n_{\max}$ is obtained.

Finally, the Lanczos coefficients $b_n$ are extrapolated beyond $n_{\max}$ according to eqs. \eqref{b_n extrapolation 1D}, \eqref{b_n extrapolation 2D}, and the autocorrelation function is calculated by numerically solving equations \eqref{coupled equations}. The latter system of equations is truncated at some large $k\gg n_{\max}$ chosen such that the result is insensitive to the precise value of $k$. In practice, we find it appropriate to choose $k=500$.

To estimate the maximal time $t_{\max}$ until which our results are reliable, we reiterate this final step with the extrapolation based on $(n_{\max}-1)$ Lanczos coefficients, and require that the discrepancy between the two approximations to $C(t)$ remains below some small $\epsilon$ ($\epsilon=10^{-3}$ for plots in Fig. \ref{fig 1}).

\medskip
\noindent{\it 1D Ising model.}~
The Hamiltonian of the model reads
\begin{equation}\label{H 1D Ising}
H=\sum\nolimits_j \sigma^x_{j}\sigma^x_{j+1}+h_z \sum\nolimits_j \sigma^z_{j}+ h_x\sum\nolimits_j \sigma^x_{j},
\end{equation}
where $\sigma^{x,y,z}_j$ are Pauli matrices at the $j$'th site and $h_x$, $h_z$ are two parameters of the Hamiltonian.
%Translation invariance is imposed by identifying $j=N+1$ with $j=1$.
This model is integrable when $h_x=0$ or $h_z=0$, and nonintegrable otherwise.
The observable we consider is the magnetization in the $z$-direction,
\begin{equation}\label{1D Ising observable}
A=\sum\nolimits_{j} \sigma^z_{j}.
\end{equation}

We are able to calculate $n_{\max}=45$   nested commutators and corresponding moments symbolically (the previous record result was $n_{\max}=38$ moments calculated numerically \cite{Noh_2021}). For example,
\begin{align}\label{1D Ising moments}
  \mu_2 & = 8+4 h_x^2, \nonumber \\
  \mu_4 & = 128+192\, h_x^2+128 \,h_z^2+16\, h_x^4+16 \,h_x^2 h_z^2.
\end{align}
%Other moments are available as part of the Supplemental Material \cite{supp}.

The corresponding Lanczos coefficients for $h_x=h_y=1$ are shown in Fig. \ref{fig 1}(a). They are consistent with the UOGH and feature pronounced odd-even alterations on top of the leading asymptote. The corresponding autocorrelation function is presented in Fig. \ref{fig 1}(d). We benchmark our result by a numerically exact computation for a finite spin chain  large enough to neglect finite size effects on the considered timescale. The long-time average $\overline C$ is nonzero, consistent with odd-even alterations of the Lanczos coefficients.

Note that, as evident from Fig. \ref{fig 1}(d), the relaxation is far from being complete up to the maximal time available. This can be attributed to an unusually long relaxation timescale of the model \eqref{H 1D Ising} \cite{Birnkammer_2022_Prethermalization} (see also \cite{wang2023diffusion,thomas2023comparing} for related observations).

%%%%%%%%%%%%%%%%%%%%%%%%%%%%%%%%%%%%%%%%%%%%%%%%%%%%%%%%%%%%%%%

\medskip
\noindent{\it 2D  $XX$-$YY$ model.}~
This is a spin-$1/2$ model on a square lattice with the  Hamiltonian
\begin{equation}\label{H 2D XX-YY}
H=\sum_{\langle \ii \jj \rangle_{-}} \sigma^x_{\ii}\sigma^x_{\jj}+ v \sum_{\langle \ii \jj \rangle_{|}} \sigma^y_{\ii}\sigma^y_{\jj}.
\end{equation}
Here $\ii$ and $\jj$ enumerate sites of the lattice, and the first (the second) sum runs over nearest neighbour sites connected by horizontal (vertical) bonds, each bond being counted once. We choose the first term of the above Hamiltonian as the observable $A$.

We manage to calculate $n_{\max}=17$ moments, with the first few given by
\begin{align}\label{2D XX-YY}
  \mu_2 & =16 \, v^2,  \nonumber \\
  \mu_4 & =640 \, v^2 (1+v^2), \nonumber \\
  \mu_6 & =2048 \, v^2 (17+39 v^2+17 v^4).
\end{align}

In Fig \ref{fig 1}(b),(e) we plot the Lanczos coefficients and the autocorrelation function for $v=1$. At this specific value of $v$ the long-time average of the autocorrelation function is fixed by symmetry to be $\overline C =1/2$. One can see that, in contrast to the previous case, $C(t)$  relaxes close to this value within the timescale accessible by our method. $C(t)$ is additionally benchmarked by the polynomial bounds~\eqref{inequality}.

%%%%%%%%%%%%%%%%%%%%%%%%%%%%%%%%%%%%%%%%%%%%%%%%%%%%%%%%%%%%%%%

\medskip
\noindent{\it 2D  Ising model.}~
The Hamiltonian is defined on a square lattice and reads
\begin{equation}\label{H 2D Ising}
H=\sum_{\langle \ii \jj \rangle} \sigma^x_{\ii}\sigma^x_{\jj}+h_z \sum_{\jj} \sigma^z_{\jj},
\end{equation}
where the first sum runs over pairs of neighbouring sites.

With an eye on computing the diffusion constant, we choose the energy current along the horizontal direction as the observable:
\begin{equation}\label{2D Ising observable}
A=J= h_z \sum_{
\substack{~\,\langle \ii \jj \rangle_{-}\\
                  \ii \prec \, \jj
        }
}
(\sigma^x_{\ii} \sigma^y_{\jj}-\sigma^x_{\jj} \sigma^y_{\ii}).
\end{equation}
Here the sum runs over horizontal bonds, the site $\ii$ being always to the left of the site $\jj$.

We are able to calculate $n_{\max}=21$   nested commutators and corresponding moments symbolically (previously 13 \cite{Heveling_2022_Numerically} and 16 \cite{de2024stochastic} moments were calculated for a different observable). First three moments read
\begin{align}\label{2D Ising moments}
  \mu_2 & =8,  \nonumber \\
  \mu_4 & =64 \, (2 + h_z^2), \nonumber \\
  \mu_6 & =1024 \,(2 + 5 h_z^2 + h_z^4).
\end{align}

The Lanczos coefficients and the autocorrelation function are shown in Fig. \ref{fig 1}(c). In contrast to previous cases, the irregularities of the Lanczos coefficients do not follow the odd-even alteration pattern. This is consistent with the fact that the autocorrelation function of the current relaxes to zero. We therefore do not include the alteration term in the extrapolation. One can see that again the autocorrelation function essentially relaxes to equilibrium within the  accessible timescale.

We further compute the diffusion constant  for a range of magnetic fields $h_z$, see Fig.~\ref{fig 2}. The convergence of the approximation \eqref{approximate C} appears to be quite good away from the integrable points $h_z=0$ and $h_z\rightarrow \infty$, as illustrated in Fig.~\ref{fig 2}(b). We conservatively estimate the uncertainty of our calculation as a maximal discrepancy between ten approximations obtained for $r$ from $(n_{\max}-9)$ to $n_{\max}$. This uncertainty is indicated in Fig.~\ref{fig 2}(a). It is below~1\% for fields~$h_z \sim 1$ but grows rapidly when $h_z$ or $h_z^{-1}$ approach zero.

%%%%%%%%%%%%%%%%%%%%%%%%%%%%%%%%%%%%%%%%%%%%%%%%%%%%%%%%%%%%%%%%%%%%%%%%%%%%%%%%%%%%%%%%%%%%%%
%%%%%%%%%%%%%%%%%%%%%%%%%%%%%%%%%%%%%%%%%%%%%%%%%%%%%%%%%%%%%%%%%%%%%%%%%%%%%%%%%%%%%%%%%%%%%%

\medskip
\noindent{\it Discussion and outlook.}~ In summary, we have advanced the recursion method to the point it can handle the dynamics of two-dimensional lattice systems over the whole relaxation timescale. We have illustrated the power of the method by computing infinite-temperature autocorrelation functions and the diffusion constant for spins $1/2$ on a square lattice.

The most resource-consuming part of our computations is performed symbolically, which means that the whole parameter space of the Hamiltonian is covered in a single run. The accuracy of the method, however, differs across the parameter space. Remarkably, the method works best deep in the nonperturbative regime, where the sequence of the Lanczos coefficients  converges to its asymptotic form most rapidly \cite{Parker_2019}.

An important ingredient of the method is the extrapolation of Lanczos coefficients beyond those explicitly computed. The extrapolation is based on the conjectured leading~\cite{Parker_2019} and subleading~\cite{Viswanath_1994_Ordering,Parker_2019,Yates_2020_Lifetime,Yates_2020_Dynamics,Dymarsky_2021_Krylov,Bhattacharjee_2022_Krylov,avdoshkin2022krylov,Camargo_2023_Krylov}  asymptotics of the coefficients. We note that the method will benefit from better theoretical understanding of the subleading terms.

%Benchmarked by rigorous polynomial bounds at small time as well as by exact diagonalization within the whole available time span for a one-dimensional model

The generalization of the method to different lattice geometries, higher spins, lattice fermions or bosons is conceptually straightforward. Finite but high temperatures can be handled by using the recursion method in conjunction with the high-temperature expansion \cite{Bhattacharyya_2024_Metallic}. Addressing lower temperatures can be more challenging, most likely necessitating a considerable amendment of the method. In particular, employing more complex scalar products \cite{Parker_2019,Dymarsky_2020_Quantum,tan2024scaling} beyond the simplest one \eqref{scalar product} may be required.

Finally, we note that recent approaches  \cite{Rakovszky_2022_Dissipation-assisted,Schuster_2023_Operator,White_2023_Effective,thomas2023comparing,ermakov2024unified} to effectively constraint the Heisenberg evolution within smaller subspaces of the operator space can potentially greatly reduce the computational cost of the method. Another very recent promising move in the same direction is a stochastic sampling of operator growth \cite{de2024stochastic}.

%%%%%%%%%%%%%%%%%%%%%%%%%%%%%%%%%%%%%%%%%%%%%%%%%%%%%%%%%%%%%%%%%%%%%%%%%%%%%%%%%%%%%%%%%%%%%%
%%%%%%%%%%%%%%%%%%%%%%%%%%%%%%%%%%%%%%%%%%%%%%%%%%%%%%%%%%%%%%%%%%%%%%%%%%%%%%%%%%%%%%%%%%%%%%

\medskip
\begin{acknowledgments}
\noindent{\it  Acknowledgments.}
OL thanks Anatoly Dymarsky and Alexander Avdoshkin for a useful discussion at the initial stage of this study. This work was supported by the Russian Science Foundation under grant \textnumero~24-22-00331, \url{https://rscf.ru/en/project/24-22-00331/}
\end{acknowledgments}

%\bibliography{confDlitra,Floquet,dynamically_integrable,LZ_and_adiabaticity}
\bibliography{C:/D/Work/QM/Bibs/open_systems,C:/D/Work/QM/Bibs/scars,C:/D/Work/QM/Bibs/recursion_method,implementation_misc}

\end{document}